\documentclass[journal]{IEEEtran}
\IEEEoverridecommandlockouts
\usepackage[utf8]{inputenc}
\usepackage[T1]{fontenc}
\usepackage{cite}
\usepackage{amsmath,amssymb,amsfonts}
\usepackage{algpseudocode}
\usepackage{algorithm}
\usepackage{graphicx}
\usepackage{tabularx}
\usepackage{textcomp}
\usepackage{xcolor}
\algnewcommand\algorithmicforeach{\textbf{for each}}
\algdef{S}[FOR]{ForEach}[1]{\algorithmicforeach\ #1\ \algorithmicdo}
\usepackage{subfigure}
\ifCLASSINFOpdf
\else
\fi

\begin{document}
\title{Co-channel Interference Management for the Next-Generation Heterogeneous Networks using Deep Leaning.}

\author{Ishtiaq Ahmad and Aftab Hussain
\thanks{Ishtiaq Ahmad is with Gomal University, D.I.Khan, Pakistan(e-mail:ishtiaqahmad@gu.edu.pk)}
\thanks{Aftab Hussain is with Gomal University, D.I.Khan, Pakistan(e-mail: ~~~)}}

\maketitle

\begin{abstract}
The connectivity of public-safety mobile users (MU) in the co-existence of a public-safety network (PSN), unmanned aerial vehicles (UAVs), and LTE-based railway networks (LRN) needs a thorough investigation. UAVs are deployed as mobile base stations (BSs) for cell-edge coverage enhancement for MU. The co-existence of heterogeneous networks gives rise to the issue of co-channel interference due to the utilization of the same frequency band. By considering both sharing and non-sharing of radio access channels (RAC), we analyze co-channel interference in the downlink system of PSN, UAV, and LRN. As the LRN control signal demands high reliability and low latency, we provide higher priority to LRN users when allocating resources from the LRN RAC shared with MUs. Moreover, UAVs are deployed at the cell edge to increase the performance of cell-edge users. Therefore, interference control techniques enable LRN, PSN, and UAVs to cohabit in a scenario of sharing RAC. By offloading more PSN UEs to the LRN or UAVs, the resource utilization of the LRN and UAVs BSs is enhanced. In this paper, we aim to adopt deep learning (DL) based on enhanced inter-cell-interference coordination (eICIC) and further enhanced ICIC (FeICIC) strategies to deal with the interference from the PSN to the LRN and UAVs. Among LRN, PSN BS, and UAVs, a DL-based coordinated multipoint (CoMP) link technique is utilized to enhance the performance of PSN MUs. Therefore, if radio access channels are shared, utilization of DL-based FeICIC and CoMP for coordinated scheduling gives the best performance.
\end{abstract}
\begin{IEEEkeywords}
Deep learning, inter-cell-interference, public-safety network, railway networks, unmanned aerial vehicles, radio access channels.
\end{IEEEkeywords}
\section{Introduction}
Emergency rescue teams (ERTs) and public safety networks (PSNs) are in charge of establishing secure surroundings and executing mission-critical responsibilities, such as acknowledging possible disasters conducted by either human or natural causes. Border patrol, emergency medical care, firefighting, police departments, and emergency policing are among the emergency relief services. PSNs play a crucial role in protecting the public, business, and the state's infrastructure during crises like fires, terrorist attacks, and ecological disasters.\cite{a1}.
\\In contrast to traditional phone services, PSN mobile users (MUs) provide voice conversations in incredibly effective ways under very rigorous guidelines. The need for network algorithms in PSNs, such as digital sharing, streaming video, and video calls, is rising as more jobs are accomplished efficiently. Giving the central unit real-time video streaming, for instance, will empower them to precisely evaluate the situation and dispatch, sufficient first responders, to handle the calamity and minimize any potential fatalities. PSNs currently use land mobile radio and conventional narrowband communication technologies, which may provide reliable voice communications but cannot support broadband transmission and often have a restricted range of compatibility and coverage \cite{a2}. To protect lives, and stop harm, criminal activity, and terrorism, emergency professionals across the nation may converse and exchange essential voice, video, and data using PSN communication. \cite{a3}.
\\The introduction of Unmanned aerial vehicles (UAVs), such as tiny drones, balloons, or gliders enables ubiquitous broadband access and possesses the significant potential to alter PSN technology \cite{a4}. UAVs have been able to find a variety of significant uses as a result of recent technical advancements, shrinking, and open-source technology efforts. \cite{a5}. Moreover, UAVs are a possible answer for civil services including emergency preparedness, sporting activities, and traffic monitoring. Furthermore, UAVs have garnered a lot of interest from both academics and businesses. Therefore, UAVs can function as airborne base stations (BSs), with a coverage area provided by aerial tiny cells, when equipped with base station hardware. It is desirable to include UAVs within the next-generation PSN because it effectively covers the edge PSN MUs during emergencies.
\\The use of UAVs as hovering BSs can significantly increase the capabilities of mobile networks \cite{a6} by utilizing line-of-sight (LoS) connections at various altitudes and the adaptability of UAV deployment at preferred places \cite{a7}. UAVs are adapted for efficient transport in UAVs communication that provides network connectivity to ground users and maximizes the flight time of UAVs. Additionally, for the secure transportation of people and valuables, BS of Long-term evolution railway networks (LRNs) generate communication with railways on the railroad and a train's crew must function effectively. As a result, the security and safety of trains depend on the reliability of LRN. 
\\ Monitoring and controlling the plethora of network components has become impossible due to the rising complexity and diversity of mobile network topologies. Therefore, there is an unmatched scientific interest in integrating adaptable intelligent machines into future mobile networks. This trend is mirrored in the creation of interconnected systems that enable machine learning techniques. ML makes it possible to systematically mine useful data from traffic and automatically identify connections that would otherwise be too complicated for human specialists to extract. Deep learning (DL), the industry standard for machine learning, has produced outstanding results in a number of fields. Researchers in networking are also starting to understand the value and power of DL and are investigating how it may be used to address issues unique to the field of mobile networking. It makes sense to use DL in 5G wireless and mobile networks. Since they are frequently compiled from many sources, presented in diverse forms, and show intricate relationships, data produced by mobile settings in particular are becoming more and more heterogeneous. As a result, a variety of particular issues become too challenging or unworkable for typical machine learning technologies.
\\Therefore, there is an immense need to develop an efficient co-channel interference strategy algorithm in order to achieve reliable MUs assignment where PSN, UAVs, and LRN are concurrently employed. The main contribution of this paper is summarized below
\begin{itemize}
\item Analyzing scenarios with and without sharing of the RAC. Here, we first examine the advantages of RAC sharing in the co-existence of three networks. We always give more priority to the LRN user when allocating resources as the railroad signal requires more stable connectivity and low latency. 
\item To cope with co-channel interference, we utilized DL-based enhanced inter-cell interference coordination (eICIC) and further eICIC (FeICIC).
\item In order to minimize interference problems for coexisting PSN, UAV, and LRN, we use collaborating and interference coordinating techniques including coordinated-scheduling coordinated multi-point (CS-CoMP).
\end{itemize}
The rest of the paper is organized as follows. In section II, we represent the related work. In section III, we demonstrate the system model to examine co-channel interference. In Section IV, we present the proposed scheme for minimizing co-channel interference using DL-based eICIC, FeICCI, and CoMP. In section V, simulation results are given to show the improved performance of the proposed technique. Finally, Section VI concludes this paper.
\section{Related Work}
Future wireless systems \cite{a9} are considered to have UAV-assisted communication because they can supplement their terrestrial architecture with flying UAV BSs. Moreover, it is necessary to consider several challenges, including system modeling for resource management and optimization, to build viable UAV-aided cellular networks \cite{a8}. The authors proposed a heterogeneous wireless network to address the communication requirements of UAVs-assisted PSN in disaster. The authors of \cite{a9} modified the optimum transportation model for UAVs to allow them to provide network connectivity to ground users while optimizing their flight time. The researchers of \cite{a10} also suggested an effective UAV BS that could coexist with a traditional network. Tor a radio link connection in a difficult-to-deploy site, UAVs can augment traditional small BSs and offer cost-effective and low-power alternatives. As an example, authors in \cite{a11} handled a combined position and user connecting issue by using UAVs rather than terrestrial BSs. The authors in \cite{a12} used the circular packing theory for the UAV deployment to get the maximum user coverage. According to the authors in \cite{a13}, UAVs are utilized for transferring power to a sensor network in an internet-of-things (IoT) network.
\\The authors in \cite{a14} presented a point-processing-based 3D UAV deployment that took advantage of UAVs with multiple antennas to connect to ground users. The authors in \cite{a15} suggested scattering UAVs with a 3D layout while keeping mmWave networks at a safe distance. Additionally, UAVs are advantageous when there is a problem in connectivity that might occasionally occur in busy places like concerts or festivals. A technological difficulty that hasn't been addressed in the earlier research \cite{a16,a17,a18} is the potential to increase power utilization in UAV-assisted cellular connections by leveraging the chances for energy collection. By extending the air time of UAVs, it may offer cellular customers service for a significantly longer time by using energy harvesting. There isn't much research in this area that looks into UAVs with energy-harvesting characteristics for wireless networking \cite{a19,a20,a21}.
\\Considering the coexistence of UAV networks with a next-generation mobile infrastructure to offer various services is a crucial topic. UAVs can establish links in a disaster-stricken region with good planning, which is the basis for their deployment with PSN in the future. Here, radio frequency planning aids in determining the number of drones needed for coverage, the best altitude to fly at, and the maximum user rates. If UAVs are to be employed in inhabited areas, determining the route loss that signals encounter is considered a crucial problem. Although, drones are intended to function as agents for quickly reconstructing communication links. However, proper channel characterization may be analytically costly and time-consuming \cite{a21,a22,a23}.
\begin{figure}
\centerline{\includegraphics[width=0.45\textwidth]{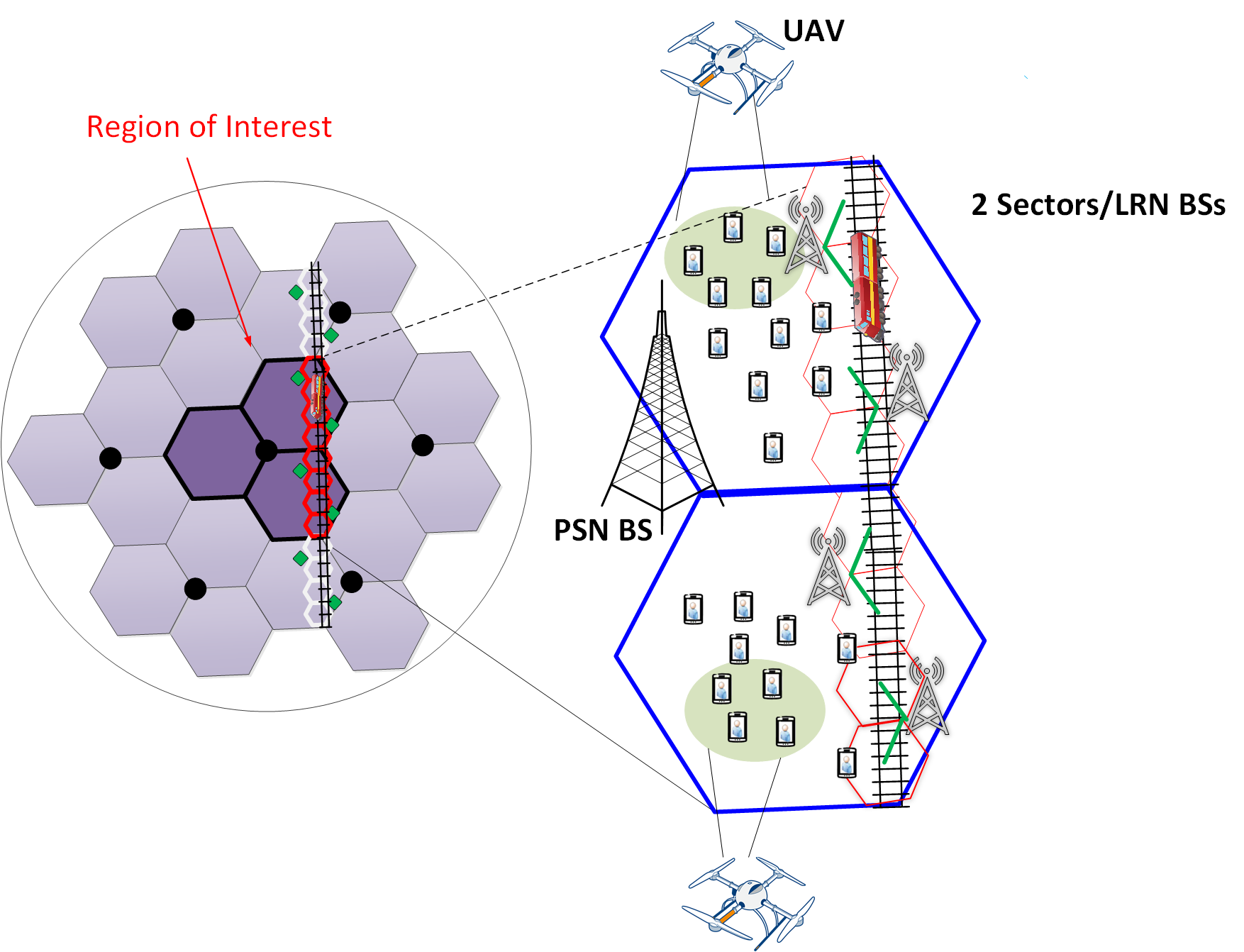}}
\caption{PSN, LRN, and UAV deployment layout.}
\label{fig1}
\end{figure}
\section{System Model}
The inter-site spacing is 4 km for PSNs and 1 km for LRN, respectively. We consider a scenario that includes a one-tier implementation of PSN BS, four LRN BSs, and two UAVs that overlap with the network's central location (center of interest), as shown in Figure \ref{fig1}. Each PSN BS has three sectors to aid transmission, whereas LRN has two sectors. Each side of the track of the railway is lined with LRN BS. Whereas UAVs are deployed at the cell edges to cover the edge users. In this study, an LTE downlink infrastructure for the co-existence of PSN-LRN and PSN-UAVs is considered. All BSs of PSN, LRN, and UAVs utilize the same system bandwidth. Physical resource blocks are assigned to each MU in the time and frequency grid.
\\Using $M$ UAVs with a $P_u$ hexagonal sector and $N$ LRN BS sites $(N = 4)$ made up of hexagonal sectors $(Q_R = 2)$, we instal a R-tier $(R = 1)$ PSN that coexists with LRN and UAV using $L$ PSN BS sites $(L = 7)$ composed of $K$ hexagonal sectors $(K_p = 3)$ within each site. Moreover, $S$ is equal to the total number of UAV, LRN, and PSN units $S = 35$. All MUs are referred to as $U$, such as PSN MU $U_p$ and LRN user $U_r$, and PSN, LRN, and UAVs BSs are referred to as $\{P-BS_1,..., L, U-BS_1,..., M, R-BS_1,..., N\}$.
\\The K-tier deployment's area of interest is the principal place for PSN, while other regions generate interference. According to Figure \ref{fig1}, the area of interest includes 2 UAVs, 4 LRN BSs, and 1 BS for PSN. MUs of PSN are scattered uniformly and at random throughout the target region for each sector. Therefore, it is likely that some UEs would be deleted close to LRN BSs. PSN MUs can be connected to LRN BSs through active wireless connection sharing. Next, we'll talk about the trade-offs involved in switching between sharing and not sharing a radio access network. The following general equation is used to determine the pathloss to each link:
\begin{equation}
    L = G_a-\zeta_l-sh-fad
\end{equation}
where $G_a$ represents the antenna gain, $\zeta_l$ is a link's pathloss, $sh$ stands for shadowing, and $fad$ presents the fading of channel. The rural macro model of the 3GPP standards is considered when assessing path loss. The macro route loss model, which was created to use a frequency range equal to 700 MHz, is stated as follows:
\begin{equation}
   \begin{split}
    \zeta_l & = 69.55 + 26.16 \log_{10} (freq)-13.82 \log_{10} (H)
    \\ & +[44.9 - 6.55 \log_{10} (H)] \log(Dist)-4.78(\log_{10} (freq))^2
    \\ & +18.33 \log_{10}(freq)-40.94
    \end{split}
\end{equation}
where $H$ is the height of the BS antenna, $Dist$ represents the distance from the BS to MUs, and $freq$ is the frequency. The shadowing caused by obstructions between MUs and BSs is modeled using a log-normal distribution and inter-site correlation coefficient equal to 0.5. The rapid signal level changes in multipath communication are characterized by fast fading. In this study, fast fading is generated for PSN MUs and high-speed train users with high mobility using the D1 and D2a models (offered by Winner II), respectively. The vertical $v$ and horizontal $h$ trims of the 3D antenna designs for the BS are calculated as follows:
\begin{equation}
    Antenna (\beta, \gamma) = - [Antenna_v(\beta)_v + Antenna_h(\beta)_h,Antenna_j]
\end{equation}
\\The practical traffic models are used to mimic typical traffic and its congestion circumstances. Depending on the employed software, including voice-over-IP (VoIP) and video, users can be classified. For this study, we assume that the modeling of traffic at the user level must be applied to the LTE standard. The following is the generic proportionately fair scheduler equation:
\begin{equation}
    Sch = \arg \max \frac{R_i}{\Bar{R_i}}
\end{equation}
where $R_i$ represents the instantaneous datarate and $\Bar{R_i}$ represents the averaged datarate. The objective is to constantly provide the LRN user with the prioritized resources first since the downlink communication of an LRN control signal necessitates more reliable connectivity and low latency.
\section{Problem Formulation}
In this paper, our main goal is to combine the eICIC, CoMP, and FeICIC to increase coverage while minimum channel $CH$ impact while staying within the constraints of data rate and average BS transmit power. Therefore, the optimization strategy is described as:
\begin{equation}
 \begin{split}   
    &\max \min_{\Delta CH} \sum_{m \in M} \alpha_m \\
    & s.t.(a): \alpha_m^{min} \leq \alpha_m \\
    & (b): \sigma\sigma^{CH} \leq \psi_{max}
 \end{split}   
\end{equation}
where the minimal throughput $\alpha_m^{min}$ of a $m-th$ MU is determined by the maximum power $(\psi_{max})$ consumption threshold and channel effects.
\section{Proposed ME-DRL Algorithm}
\section{Proposed DL-based Interference Control Technique}
As shown in Figure \ref{fig1}, we use cooperative communication solutions, such as coordinated scheduling CoMP using eICIC and FeICIC, to evaluate interference for the present public safety, UAV, and railway networks. In this respect, we classify the UAV, railway, and public safety systems into various circumstances and assess each one's efficacy. Below are descriptions and details of the situations.
\subsection{Situation 1: UAVs, LRN, and PSN Without RAC Sharing}
PSN UEs cannot access LRN BS or UAV, whose coverage is less similar to that of the public safety network. It is similar to macrocells and femtocells that may coexist with closed subscriber subgroups when the low-power nodes only allow a certain number of UEs to access the network. If LRN/UAV BS is located close to the PSN BS, there is a critical co-channel interference. It is because of the limited coverage of LRN/UAV BS that have high-power nodes along the track. For the PSN, MU is also in the LRN/UAV network's center coverage and is receiving a relatively low-power intended signal. Moreover, the interference intensity from LRN/UAV BS is high for the UE towards the edge of PSN coverage.
\\The power setting for eICIC/FeICIC schemes is designed to shield the PSN MUs located near the closed subscriber group LRN/UAVs by minimizing the transmit power of the LRN/UAVs. As LRN/UAVs BSs are low-power access points and are generally deployed by users for limited service standards, offering the small cells lower priority. Therefore, reducing LRN/UAVs BS transmission power makes it logical to protect PSN MUs. However, operators have set up high-power nodes for the UAV, LRN, and PSN. In a case where radio access network sharing is not allowed, reducing the power of LRN BSs will also harm the network's reliability and cause service disruptions, rendering the eICIC/FeICIC schemes useless.
\subsection{Situation 2: UAVs, LRN, and PSN With RAC Sharing}
Instead of being sources of excessive interference, LRN BS/UAVs might be viewed as BSs to increase coverage at the cell boundaries of the PSN through active RAC sharing. In a RAC sharing environment, PSN MUs can attach to LRN BSs/UAVs, which reduces co-channel interference and increases LRN BS/UAVs resource consumption. Since the LRN UE travels along the track and generally receives greater power from LRN BSs, the PSN does not need to provide RAC sharing for LRN users. This paper considers sharing the LRN/UAV RAC by PSN UEs.
\\In the RAC sharing scenario, we considered predefined scheduling constraints. The LRN UE must receive the finest resources since its downlink transmission requires minimal latency and high reliability. As a result, we give LRN UE higher priority when allocating resources to meet the needs of the LRN's mission-critical service. While PSN BSs use proportionate fair scheduling to schedule their MUs.
\subsection{Situation 3: FeICIC/eICIC UAVs, LRN and PSN Without RAC Sharing with Coordinated Scheduling-CoMP}
In this case, we offer effective interference mitigation solutions that will improve the reliability and throughput of the PSN MU performance and the quality of the LRN user link. It should be noted that the co-existence of UAV, LRN, and PSN is a novel heterogeneous network with overlapping PSN. Given the difference in cell sizes and the ability of LRN BSs/UAVs to unload PSN MUs via RAC sharing, this situation is comparable to the heterogeneous network configuration of macro and pico cells. Here MUs can connect with all kinds of BSs, and pico BSs with lower coverage are destined to unload macro UEs.
\\Although there is a main difference between the two situations, PSN MUs may be found everywhere and are regarded as normal users. In contrast, LRN users move along a track that is positioned among LRN BSs and UAVs stationed at the cell edges. To target this situation with RAC sharing and minimize the interference from PSN BSs to the LRN/UAV network, time-domain eICIC and FeICIC approaches are employed. The surrounding LRN BSs often cause more interference to the LRN user present along the track than the PSN BSs. To lessen this interference, the nearby LRN BSs may explore coordinated scheduling CoMP.
\\To avoid interference from the LRN/UAVs BSs, no data is sent from PSN BS during absolute blank subframes (ABSs) for eICIC. Moreover, FeICIC is used to facilitate data transmission for center PSN MUs during power-reduced ABSs, with moderate user interference from LRN BSs and UAVs. However, the sole beneficiaries of the deployment of ABS/reduced power-ABS will be cell-edge users of LRN BSs and UAVs. Cell range extension is used to avoid interference for the PSN MUs, which are subject to severe interference from LRN/UAVs BSs. By applying a heavy bias to received signal-reduced power of LRN/UAVs BSs to offload, so they subsequently become beneficial from ABSs/reduced power-ABSs.
\\When PSN BSs are under too much strain, the network coverage of LRN/UAVs may be increased to allow more PSN MUs to be offloaded, and the remaining PSN MUs with better channel quality can then be well supported by PSN BSs. In such a situation, the offloaded UEs will experience considerable co-channel interference from the PSN BSs. To lessen interference to the dispersed MUs, ABS/reduced power-ABS is employed.
\\In this non-RAC sharing instance, we also consider coordinated scheduling CoMP for the PSN, UAVs, and LRN BSs. In order to avoid interference with the demands of the LRN user's mission-critical service, the aggressor BSs muffle their physical resource blocks with the assistance of the CoMP sites. However, LRN customers are given a higher priority when resources are allocated. Coordinated scheduling CoMP is used between PSN and LRN BSs, PSN and PSN BSs, PSN and UAVs, UAVs and UAVs BSs, and LRN and LRN BSs.
\subsection{Situation 4: DL-based FeICIC/eICIC UAVs, LRN, and PSN Without RAC Sharing with Coordinated Scheduling CoMP}
\begin{figure}
\centerline{\includegraphics[width=0.45\textwidth]{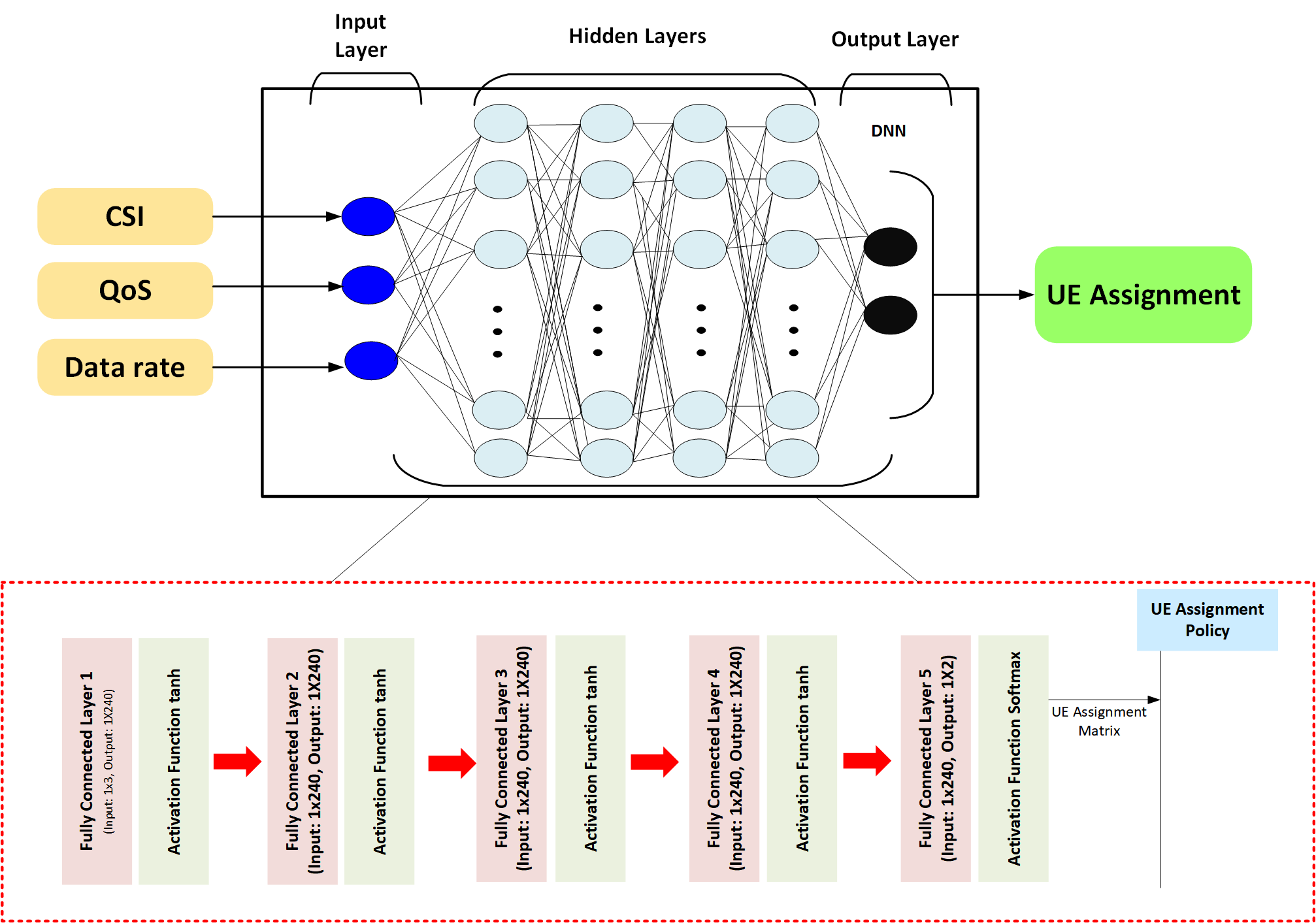}}
\caption{DL-based user assignment to PSN, LRN, and UAV.}
\label{fig2}
\end{figure}

A DL is necessary to satisfy QoS and data rate requirements. Input nodes for channel status information (CSI), maximum throughput, hidden nodes, and output units for eICIC/FeICIC are required for this DL, along with coordinated scheduling (CoMP) and output units for eICIC/FeICIC. For convenience, both the hidden layers and the outputs layer are referred to as layers. A special input characteristic of the recommended DL structure is produced by combining the QoS, CSI, and maximum throughput, as shown in Figure \ref{fig2}. The proposed DL technique uses any feed-forward architecture, such as fully linked layers. We attempt to approximate the uncertain radio access network sharing computing technique using the DNN.
\\By consciously allocating users to BS, the DNN creates an output vector. The output layer activation must be carefully designed to consistently deliver an adequate solution that fulfills the power restriction shown in (a) and the efficiency constraint shown in (b). As there are more BS antennas and users, there are more factors to anticipate. Therefore, the technique should avoid the overfitting issue with subpar performance. We examine a DNN that creates low-dimensional transitional variables that are crucial elements in creating the optimum beamforming solution to address the overfitting problem. We construct the DNN's output layer, which changes depending on the training sample. We start by using a scaling softmax method to get a usable user matrix. After learning, we then get the user's assignment matrix. By using softmax as the final activation function, the DNN then transfers the resultant allocation.
\\We considered three inputs—QoS, Throughput, and CSI—and sent the input data in order to fully link the hidden network. The 1st layer of the hidden system, which had four fully connected layers and generated 240 values from three input matrices, utilized the tangent hyperbolic non-linear activation The 1st, 2nd, and 3rd layers employ a non-linear activation (tangent hyperbolic), whereas the fourth and final levels employ softmax. Additionally, the output layer produces approximate UE assignment matrices.
\begin{figure}
\centerline{\includegraphics[width=0.45\textwidth]{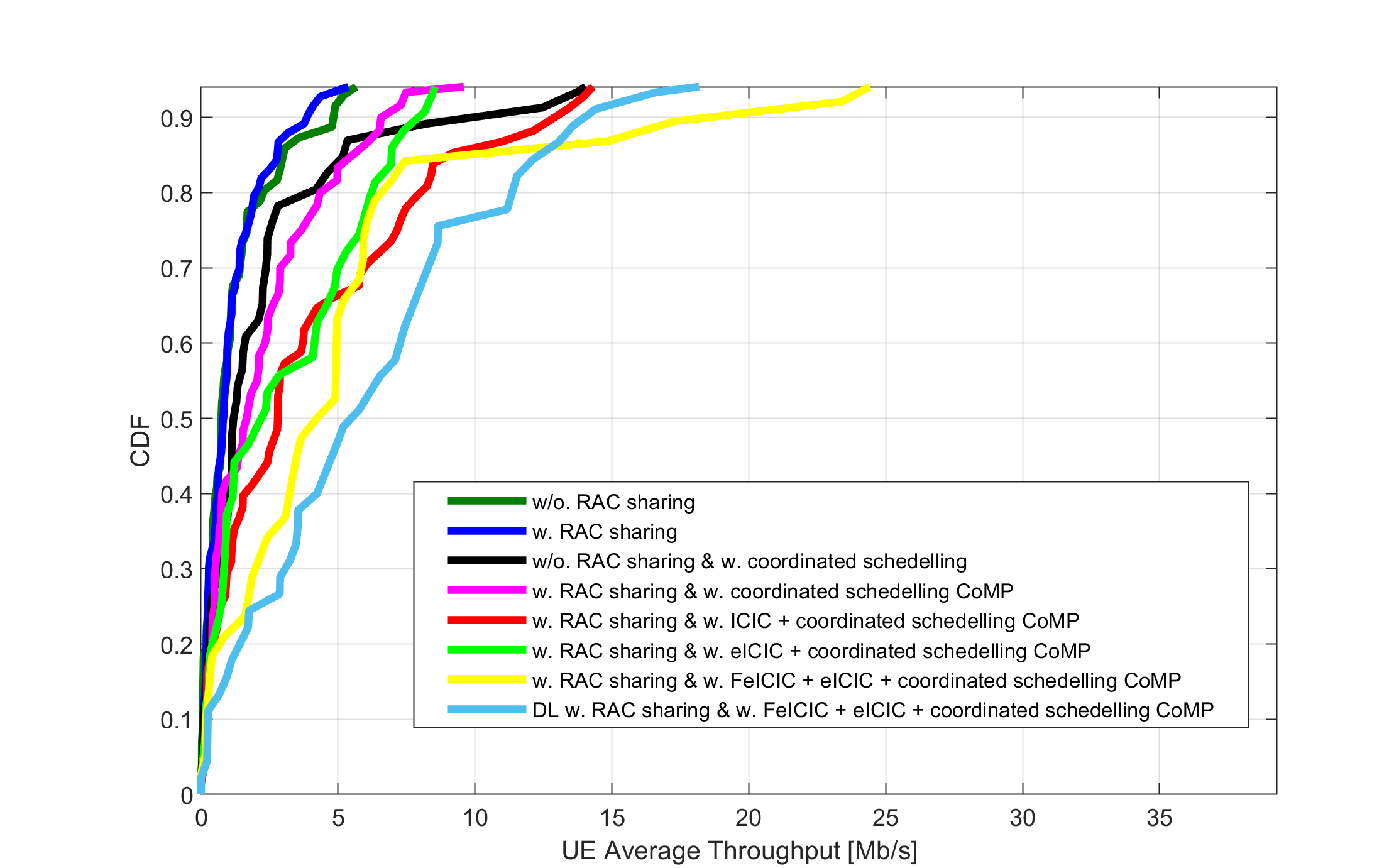}}
\caption{User's Average Throughput.}
\label{fig3}
\end{figure}
\section{Simulation Results}
This section evaluates the proposed scheme performed with its communication situations in various co-existing LRN and PSN scenarios. System-level simulations (SLS) are carried out using the one-tier PSN coexistence with an LRN/UAV network in order to validate the analysis of co-channel interference under various scenarios. Additionally, this work employs practical traffic models rather than taking into account the complete buffer scenario. For public safety users, we consider two forms of communication: video is 20\%, VoIP is 80\%, and the transmission of LRN control signals is assumed to be VoIP traffic. 
\\Users of PSNs are dispersed throughout the region of concern in an equitable and random manner. Even though just one sort of LRN user is anticipated, it's possible that only a small number of PSN UEs will enter the LRN BS area. This section compares the simulation outcomes for a number of concurrent PSN, UAVS, and LRN system scenarios. We compare instances of RAC sharing with instances of non-RAC sharing in order to demonstrate the advantages while adopting RAC sharing among three networks. The effectiveness of the scenarios involving an LRN network coexistence with a PSN is evaluated.
\\The efficiency of the UE throughput at 50\% of the total of the complete distributing mechanism is compared in Figure \ref{fig3} under the conditions of CDF. The comparison in Figure \ref{fig3} demonstrates unequivocally that at 50\% of CDF, the sharing RAC scenario outperforms the non-sharing one. Out of all the potential outcomes, a coordinated scheduling CoMP with RAC sharing, eICIC, and FeICIC is the best-case scenario. When we attempt to use eICIC and FeICIC, user throughput performance is greater owing to the benefit of the RAC sharing but is almost comparable to a certain throughput performance according to interference deficiencies. Situations with eICIC, FeICIC, and CoMP, therefore, show the result with very little variance. A few of these situations are affected by interference. However, the DL-based with coordinated scheduling CoMP with radio access network sharing outperforms all other situations, as can be seen in the UE throughput graph.
\\Accordingly, when user throughput is 5 Mb/sec, the CDF for MUs with RAC sharing, MUs without RAC sharing, MUs without RAC sharing and with coordinated scheduling CoMP, MUs with RAC sharing and with coordinated scheduling CoMP, MUs with RAC sharing with coordinated scheduling CoMP and with ICIC, MUs with RAC sharing with coordinated scheduling CoMP and with eICIC, MUs with RAC sharing with FeICIC, eICIC and coordinated scheduling CoMP and MUs with RAC sharing DL with FeICIC, eICIC and coordinated scheduling CoMP is 0.98, 0.9, 0.82, 0.814, 0.7, 0.67, 0.59, and 0.48, respectively. 
\begin{figure}
\centerline{\includegraphics[width=0.45\textwidth]{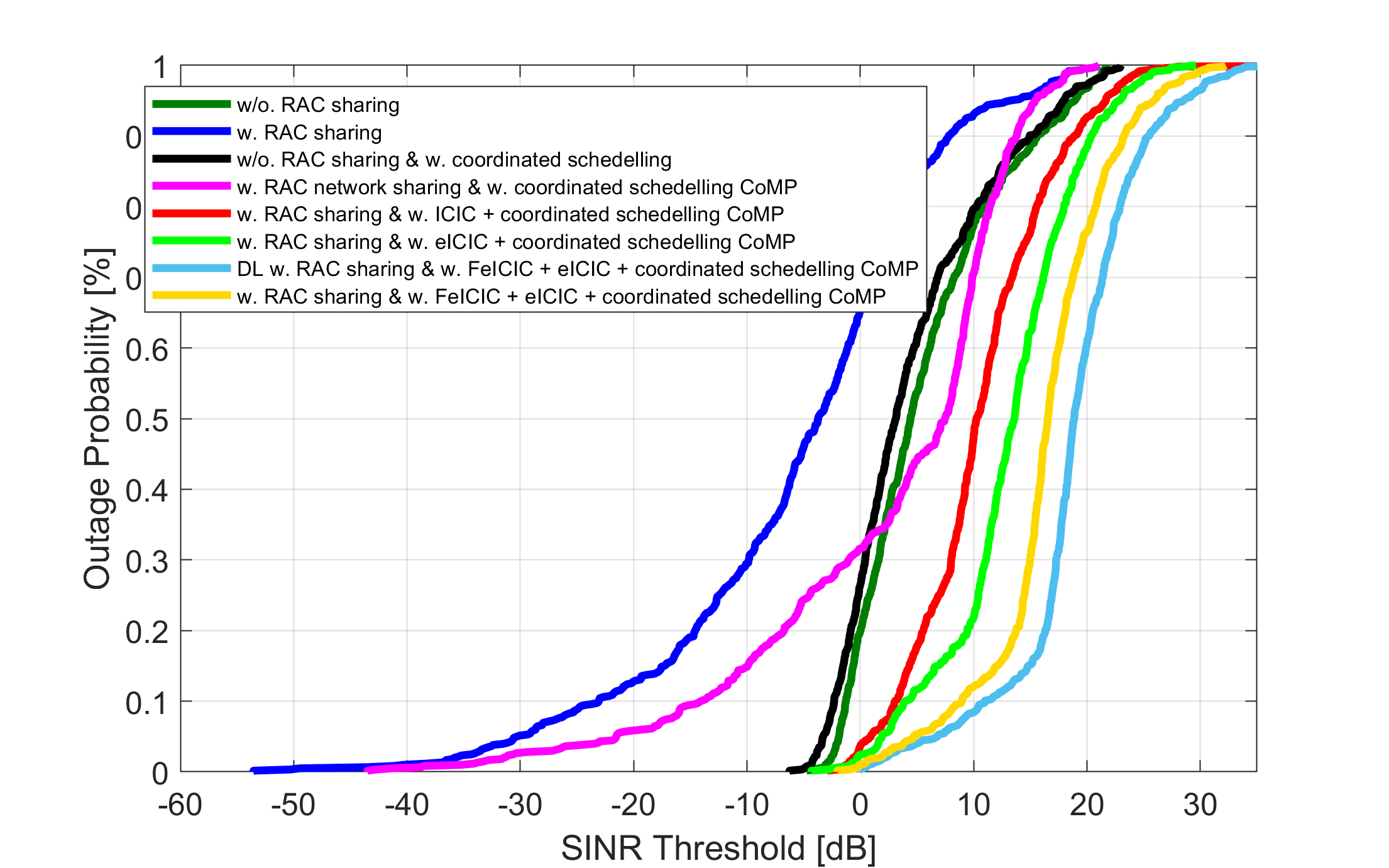}}
\caption{Outage Probability.}
\label{fig4}
\end{figure}
\\ Figure \ref{fig4} displays a plot of the SINR threshold and outage probability. The proposed DL technique outperforms other instances. The SINR values are lower in the absence of RAC sharing than in the presence of it because there is less interference. However, the DL-based with coordinated scheduling CoMP with radio access network sharing outperforms all other situations, as can be seen in the UE throughput graph. Accordingly, when the SNR threshold is 0 dB, the CDF for MUs with RAC sharing, MUs without RAC sharing, MUs without RAC sharing and with coordinated scheduling CoMP, MUs with RAC sharing, and with coordinated scheduling CoMP, MUs with RAC sharing with coordinated scheduling CoMP and with ICIC, MUs with RAC sharing with coordinated scheduling CoMP and with eICIC, MUs with RAC sharing with FeICIC, eICIC and coordinated scheduling CoMP and MUs with RAC sharing DL with FeICIC, eICIC and coordinated scheduling CoMP is 0.2, 0.66, 0.3, 0.35, 0.07, 0.06, 0.05, and 0.1, respectively.
\\The SINR values are lower when there is network sharing than when there is RAC sharing. However, the DL-based with coordinated scheduling CoMP with RAC sharing outperforms all other situations, as can be seen in the UE throughput graph. Accordingly, where -70 dBm is the interference threshold, when the SNR threshold is 0 dB, the CDF for MUs with RAC sharing, MUs without RAC sharing, MUs without RAC sharing and with coordinated scheduling CoMP, MUs with RAC sharing and with coordinated scheduling CoMP, MUs with RAC sharing with coordinated scheduling CoMP and with ICIC, MUs with RAC sharing with coordinated scheduling CoMP and with eICIC, MUs with RAC sharing with FeICIC, eICIC and coordinated scheduling CoMP and MUs with RAC sharing DL with FeICIC, eICIC and coordinated scheduling CoMP is 0.98, 0.486, 0.6, 0.35, 0.67, 0.76, 0.79, and 0.85, respectively as shown in Figure \ref{fig5}.
\begin{figure}
\centerline{\includegraphics[width=0.45\textwidth]{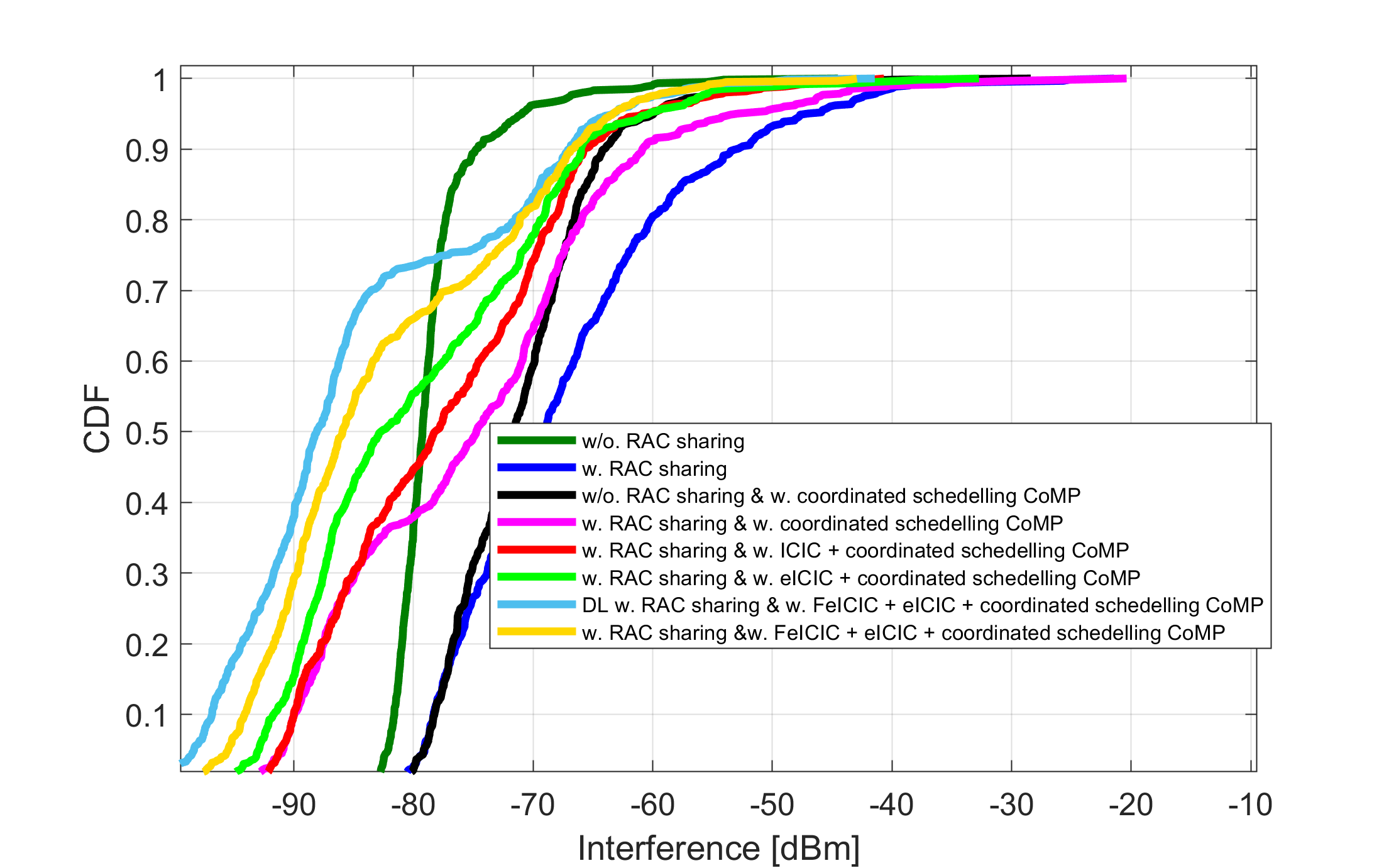}}
\caption{Interference.}
\label{fig5}
\end{figure}
\section{Conclusions}
This study uses cooperative interference coordinating approaches to demonstrate co-channel interference assessment for the coexisting PSN and LRN. To the best of our knowledge, this study of co-channel interference assessment for UAV, LRN, and PSN, is the first research to address resource allocation for coexisting LRN, UAV, and PSN. RAC sharing is used to produce high capacity and improved channel conditions when three networks coexist. The basic explanation for this improvement is rather straightforward: another BS does not allocate the identically same resources that have already been dedicated to the PSN MU with the help  of CoMP.  By developing collaboration based on user QoS preferences, it makes the greatest use of the available spectrum resources. The DL-based UE allocation is also carried out in line with the users' traffic requirements. The main aim is to find out if deploying UAVs may aid in achieving more difficult goals. Utilizing RAC sharing and coordinated scheduling CoMP, dynamic ICIC, eICIC, and FeICIC technologies, the data rate requirements of a 6G network may be met while keeping the power consumption of the wireless infrastructure within bearable bounds. Additionally, to meet the increasing expectations of the consumers, we determined the most recent advances and the areas that need uniformity. Future research and cutting-edge technology will also significantly alter how people and PSN agencies carry out their everyday operations.
\ifCLASSOPTIONcaptionsoff
  \newpage
\fi

\end{document}